\documentclass[epsfig,12pt]{article}
\usepackage{makeidx}
\usepackage{amsmath}
\usepackage{amsfonts}
\usepackage{amssymb}
\usepackage{graphicx}

\input epsf.sty
\newtheorem{theorem}{Theorem}
\newtheorem{acknowledgement}[theorem]{Acknowledgement}

\makeatletter \@addtoreset{equation}{section}
\renewcommand{\theequation}{\thesection.\arabic{equation}}
\input{tcilatex}
\begin{document}

\title{\vspace{-2cm}\rightline{\mbox{\small {\bf LPHE-MS-11-06  /
CPM-11-06}}} \bigskip \bigskip \textbf{Graphene, its Homologues and Their
Classification}}
\author{L.B Drissi$^{a}$, E.H Saidi$^{a,b,c}$ \\
{\small a. MAScIR-INANOTECH, Institute of Nanomaterials and Nanotechnology,
Rabat, Morocco,}\\
{\small b. LPHE, Modelisation et Simulation, Facult\'{e} des Sciences Rabat,
Morocco}\\
{\small c. Centre of Physcs and Mathematics, CPM-CNESTEN, Morocco}}
\maketitle

\begin{abstract}
Using tight binding model, lattice QFT and group theory methods, we study a
class of lattice QFT models that are cousins of graphene; and which are
classified by finite dimensional ADE Lie groups containing the usual
crystallographic symmetries as discrete subgroups. We show in particular
that the electronic properties of the 1D lattice poly-acetylene chain are
given by a $SU\left( 2\right) $ model and those of the well known 2D
graphene by $SU\left( 3\right) $. We also give two other models classified
by $SU\left( 4\right) $ and $SO\left( 6\right) $ symmetries; they
respectively describe 3D diamond and 3D\ lattice with octahedral sites.{\ It
is shown as well that the dispersion energies of this set of models are
completely characterized by the roots of the Lie algebras underlying the
symmetry groups. Other features, such as }${SO}\left( {5}\right) ${\ lattice
involving }${sp}^{3}{d}${\ hybridization as well as the relation between the
4D hyperdiamond, having }a $SU\left( 5\right) $ symmetry and the 4D lattice
QCD{, are also discussed.}\newline
\end{abstract}


\section{Introduction}

Tight binding model is a simple lattice quantum field theory modeling the
couplings between pairs of quantum states living at closed neighboring sites
of the crystal \textrm{\cite{1,2}}. These short range pairings have been
shown to describe quite adequately the electronic properties of graphene and
homologues \textrm{\cite{3,4,5}}. Tight binding approach is nicely
represented in QFT in terms of hops of the delocalized electrons/holes from
sites $\mathbf{r}_{\mathbf{n}}$ of the crystal to nearest neighbors at $%
\mathbf{r}_{\mathbf{n}}+\mathbf{v}_{l}$. These electron and hole hops will
be interpreted in this study in terms of step operators of Lie algebras that
appear as hidden symmetries of lattices. This remarkable feature opens a
window between continuous group representation theory, often used in
elementary particles physics, and electronic properties in solid state
physics involving representations of discrete symmetries \textrm{\cite{DS}}.
As Lie groups are not common in solid state physics; we will refer below to
these continuous groups as \emph{hidden symmetries}; the well known
crystallographic symmetries appear here as discrete subgroups of these
continuous groups.\newline
In this paper, we use tight binding method to engineer a series of \emph{N-
dimensional} lattice QFT models that are classified by ADE Lie groups of
Cartan. In this modeling, the interactions between the first nearest
neighbors are described by the basic representations of the leading elements
of the Lie groups \textrm{\cite{6,7,8}}. We show amongst others that the
dispersion energies are completely determined by the root system of the
underlying Lie algebras of these groups.\newline
Our construction gives also a Lie group representation theory explanation of
the idea of treating 2D honeycomb and higher dimensional homologue $\mathcal{%
L}$ as the superposition of two sublattices $\mathcal{A}$ and $\mathcal{B}$.
At first sight the way of thinking about such lattices $\mathcal{L}$ as $%
\mathcal{A}\cup \mathcal{B}$ seems to be a beautiful trick; but it happens
that it has a deep mathematical reason. In the case of $SU\left( N\right) $
models for instance, the two sublattices, denoted $\mathcal{A}_{SU\left(
N\right) }$ and $\mathcal{B}_{SU\left( N\right) }$, are in fact intimately
related with the \emph{fundamental} $\mathbf{\text{\b{N}}}$\ and \emph{%
anti-fundamental} \textbf{\={N}}\ representations of $SU\left( N\right) $
whose weight vectors $\mathbf{\mu }_{l}$ and $\mathbf{\mu }_{l}^{\prime }$
satisfy the following constraint relations%
\begin{equation*}
\begin{tabular}{lllll}
$\mathbf{\text{\b{N}}}$ & : & $\mathbf{\mu }_{1}+\mathbf{\mu }_{2}+\mathbf{%
\ldots }+\mathbf{\mu }_{N-1}+\mathbf{\mu }_{0}$ & $=0$ &  \\
\textbf{\={N}} & : & $\mathbf{\mu }_{1}^{\prime }+\mathbf{\mu }_{2}^{\prime
}+\mathbf{\ldots }+\mathbf{\mu }_{N-1}^{\prime }+\mathbf{\mu }_{0}^{\prime }$
& $=0$ &  \\
&  &  &  &
\end{tabular}%
\end{equation*}%
which extend the well known one $\mathbf{v}_{1}+\mathbf{v}_{2}+\mathbf{v}%
_{3}=\mathbf{0}$ of 2D graphene giving the \emph{first nearest neighbors}
and playing a central role in the study of electronic properties; in
particular dispersion energy relation and the link with Dirac relativistic
theory in 3D space time. The above constraint relations capture just the
traceless condition of $SU\left( N\right) $; and have a physical
interpretation in terms of conservation of total momenta. \newline
Moreover, the \emph{second nearest neighbors} of a lattice site at $\mathbf{r%
}_{\mathbf{n}}$, and which contribute as first order corrections in the
tight binding approach, are located at $\mathbf{r}_{\mathbf{n}}+\mathbf{v}%
_{ij}$ with the $\mathbf{v}_{ij}$'s proportional to the vectors $\mathbf{%
\alpha }_{ij}=\mathbf{\mu }_{i}-\mathbf{\mu }_{j}$ which are exactly the $%
N\left( N-1\right) $ roots of $SU\left( N\right) $ symmetry; showing,
amongst others, that the sublattices $\mathcal{A}_{SU\left( N\right) }$ and $%
\mathcal{B}_{SU\left( N\right) }$ has much to do with the root lattice of
the $SU\left( N\right) $ group. In other words, sites $\mathbf{r}_{\mathbf{n}%
}$ in the $\mathcal{A}_{SU\left( N\right) }$ and $\mathcal{B}_{SU\left(
N\right) }$ sublattices of $\mathcal{L}_{SU\left( N\right) }$ are generated
by simple roots like
\begin{eqnarray}
\mathbf{r}_{\mathbf{n}} &\sim &n_{1}\mathbf{\alpha }_{1}+n_{2}\mathbf{\alpha
}_{2}+\ldots +n_{N-1}\mathbf{\alpha }_{N-1}  \label{R} \\
&&  \notag
\end{eqnarray}%
where the integral vector $\mathbf{r}_{\mathbf{n}}=\left( n_{1}\mathbf{,}%
n_{2}\mathbf{,}\ldots n_{N-1}\right) $ and where $\left\{ \mathbf{\alpha }%
_{1},\mathbf{\alpha }_{2},\ldots ,\mathbf{\alpha }_{N-1}\right\} $ are the
\emph{simple} roots of $SU\left( N\right) $. \newline
Furthermore, by using known results on group theory methods in physics, the
standard tight binding hamiltonian itself has as well a nice group
theoretical interpretation; since the operators describing the electrons
hops turn out to be nothing but step operators of the Lie algebra of $%
SU\left( N\right) $ symmetry. Within this view, one can use $SU\left(
N\right) $ symmetry to build generalization of tight binding model that
describe higher order couplings. In this regards, and though beyond the
scope of this study, lessons learnt from \emph{2D} conformal field theories
show that one may borrow techniques from Kac-Moody algebras like Suggawara
method \textrm{\cite{7} }to extend standard tight binding method, describing
pair couplings, to implement higher order interactions. In the present
paper, we will mainly focus on exhibiting some basic features of the lattice
$\mathcal{L}_{SU\left( N\right) }$ on which live the physics of the tight
binding model. We also consider the example of $SO\left( 6\right) $ model
which has octahedral sites that form a vector representation of $SO\left(
6\right) $. \newline
To illustrate our idea, we study four examples of lattice models in diverse
dimensions \emph{D}; two of them, having respectively \emph{D=1} and \emph{%
D=2}, concern the electronic properties of the two following :\newline
(\textbf{1}) the poly-acetylene chain \textrm{\cite{9}} which, in our
classification, turns out to correspond to a $SU\left( 2\right) $ model.
Here $SU\left( 2\right) $ is the usual isospin group; it is the first
element of the $SU\left( N\right) $ series with $N\geq 1$; its basic
representation is the doublet with isospin states $|\pm \frac{1}{2}>$
satisfying the traceless property
\begin{equation*}
s_{z}^{\uparrow }+s_{z}^{\downarrow }=\frac{1}{2}-\frac{1}{2}=0.
\end{equation*}%
(\textbf{2}) graphene, a sheet of graphite classified as a $SU\left(
3\right) $ model. The basic representation of this group has three states
with quantum numbers $\mathbf{\mu }_{1},$ $\mathbf{\mu }_{2},$ $\mathbf{\mu }%
_{3}$ (weight vectors in group theory language) that should be imagined as
the extension of the weights $\pm \frac{1}{2}$ of $SU\left( 2\right) $.
These are \emph{2D} vectors that satisfy the property
\begin{equation*}
\mathbf{\mu }_{1}+\mathbf{\mu }_{2}+\mathbf{\mu }_{3}=\mathbf{0}.
\end{equation*}%
Two others \emph{3D} lattice QFT systems given by the $SU\left( 4\right) $
and $SO\left( 6\right) $ models respectively based on tetrahedral (diamond)
and octahedral crystals. To have delocalized electrons described by tight
binding approach on these 3D lattices, one has to go beyond the usual $sp^{n}
$ hybridizations of carbon atoms; one needs for example material alloys with
atoms having $sp^{3}d^{1}$\ ($sp^{3}d^{3}$) hybridizations with delocalized
electrons hoping to first nearest neighboring tetrahedral (octahedral)
sites. But to illustrate the general idea in simple words, we consider
rather two toy models for diamond and octahedron respectively based on the
crystals\textrm{\ }$\mathcal{L}_{SU\left( 4\right) }$\textrm{\ and }$%
\mathcal{L}_{SO\left( 6\right) }$ described in section 4 and 5\textrm{.}%
\newline
The presentation is as follows: In section 2, we study the electronic
properties of the ideal poly-acetylene chain. In section 3, we consider the
case of graphene and show that it is precisely classified by the $SU\left(
3\right) $ group. In section 4, we develop the $SU\left( 4\right) $ diamond
and in section 5, we study the octahedral model based on $SO\left( 6\right) $%
. Last section is devoted to conclusion and comments.

\section{Poly-acetylene chain as a $SU\left( 2\right) $ model}

Roughly, the poly-acetylene\textrm{\footnote{%
The acetylene is $C_{2}H_{2}$ with carbons in the $sp^{1}$ hybridization
with \emph{2} delocalized electrons. By chain we mean the $C_{m}H_{2}$
generalization with large integer m.}} chain is a linear molecule of carbon
atoms in the $sp^{1}$ hybridization with delocalized electrons ($2$
pi-electrons per carbon atom). To study the electronic properties of this
organic molecule, which in the present paper we take it in the ideal case;
that is an infinite chain, we need to specify two main things:

\begin{itemize}
\item the 1-dimensional lattice of the carbon chain where live the
delocalized electrons,

\item the tight binding hamiltonian $\mathcal{H}_{{\small SU}\left( {\small 2%
}\right) }$ describing the couplings between electrons belonging to closest
neighboring atoms.
\end{itemize}

\begin{figure}[tbph]
\begin{center}
\hspace{0cm} \includegraphics[width=10cm]{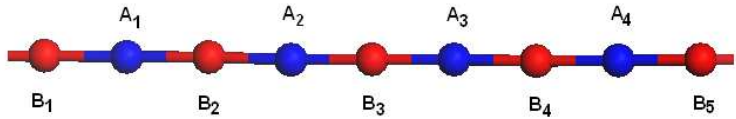}
\end{center}
\par
\vspace{-0.5 cm}
\caption{{\protect\small Lattice }$\mathcal{L}_{su\left( 2\right) }$%
{\protect\small \ given by the superposition of two sublattices }$\mathcal{A}%
_{su\left( 2\right) }${\protect\small \ and }$\mathcal{B}_{su\left( 2\right)
}${\protect\small . }}
\label{1}
\end{figure}

\emph{the lattice}\newline
The \emph{1D} lattice, denoted as $\mathcal{L}_{su\left( 2\right) },$ is
depicted in fig(\ref{1}); it is isomorphic to the one- dimensional integer $%
\mathbb{Z}$- lattice with coordinates $x_{m}=md$ where $d$ is the length of
the carbon-carbon bond. There are two different, but equivalent ways, to
deal with this lattice; one of them is that based on mimicking the study of
the\emph{\ 2D} honeycomb of graphene; it is given by the superposition of
two sublattices $\mathcal{A}_{su\left( 2\right) }$ and $\mathcal{B}%
_{su\left( 2\right) }$ (blue and red in the figure).

\ \ \ \newline
Each atom of this lattice, say an atom $A_{n}$ of sublattice $\mathcal{A}%
_{su\left( 2\right) }$, has two first nearest neighbors of B-type; that is $%
B_{n}$ and $B_{n+1}$; for illustration see fig(\ref{1}). The quantum states
of the delocalized electrons of these atoms are described by the wave
functions%
\begin{equation}
\begin{tabular}{lll}
$A_{n}\left( x\right) $ & $=\dint_{-\infty }^{+\infty }\frac{dk}{2\pi }%
e^{ikx}\tilde{A}_{n}\left( k\right) $ & , \\
&  &  \\
$B_{n}\left( x+v_{i}\right) $ & $=\dint_{-\infty }^{+\infty }\frac{dk}{2\pi }%
e^{ik\left( x+v_{i}\right) }\tilde{B}_{n}\left( k\right) $ & ,%
\end{tabular}%
\end{equation}%
where $v_{0}$ and $v_{1}$ are the relative positions parameterizing the two
first nearest neighbors. These $\mathbf{v}_{i}$'s satisfy the remarkable
constraint relation%
\begin{equation}
\begin{tabular}{llll}
$v_{0}+v_{1}=0$, & $\rightarrow $ & $v_{0}=-v_{1}=v$ & ,%
\end{tabular}
\label{BT2}
\end{equation}%
that turns out to have a nice group theoretic interpretation. More
precisely, $v_{0}$ and $v_{1}$ are proportional to the weights of the
isospinorial representations of $SU\left( 2\right) $. By setting $%
v_{1}=2d\mu _{1}$ and $v_{0}=2d\mu _{0}$, it is clear that the constraint eq(%
\ref{BT2}) is solved by the fundamental weights $\mu _{0}=+\frac{1}{2}$ and $%
\mu _{1}=-\frac{1}{2}$ of the $SU\left( 2\right) $ doublet representation
with dominant weight $\mu _{0}$ and trace
\begin{equation}
\begin{tabular}{ll}
$2d\left( \mu _{0}+\mu _{1}\right) =2d\left( \frac{1}{2}-\frac{1}{2}\right)
=0$ & .%
\end{tabular}%
\end{equation}

\emph{the hamiltonian}\newline
By focusing on correlations between the first nearest neighbors, the tight
binding hamiltonian reads as follows
\begin{equation}
\begin{tabular}{lll}
$\mathcal{H}_{SU\left( 2\right) }$ & $=-t\left( \mathcal{F}_{v}+\mathcal{F}%
_{v}^{\dagger }\right) -t\left( \mathcal{G}_{v}+\mathcal{G}_{v}^{\dagger
}\right) \ $ & ,%
\end{tabular}
\label{23}
\end{equation}%
with $t$ is the hop energy and where $\mathcal{F}_{v}$ and $\mathcal{G}_{v}$
are operators realized in terms of the electronic creation and annihilation
ones $A_{x_{m}}^{\pm },$ $B_{x_{m}}^{\pm }$ as follows,%
\begin{equation}
\begin{tabular}{llll}
$\mathcal{F}_{v}=\dsum\limits_{m\in \mathbb{Z}}A_{x_{m}}^{-}B_{x_{m}+v}^{+}$
& , & $\mathcal{F}_{v}^{\dagger }=\dsum\limits_{m\in \mathbb{Z}%
}A_{x_{m}}^{+}B_{x_{m}+v}^{-}$ & , \\
$\mathcal{G}_{v}=\dsum\limits_{m\in \mathbb{Z}}A_{x_{m}}^{-}B_{x_{m}-v}^{+}$
& , & $\mathcal{G}_{v}^{\dagger }=\dsum\limits_{m\in \mathbb{Z}%
}A_{x_{m}}^{+}B_{x_{m}-v}^{-}$ & . \\
&  &  &
\end{tabular}
\label{FG}
\end{equation}%
Notice by the way that the operator $\mathcal{F}_{v}$ describes electronic
hops from left to right while $\mathcal{G}_{v}$ describes hops from right to
left. These two hops generate two kinds of symmetries that we want to
comment through a set of remarkable properties that turn out to be also
valid for the higher dimensional extensions to be considered in next
sections.

\emph{property 1}\textbf{:} the above hamiltonian $\mathcal{H}_{SU\left(
2\right) }$ is an adequate approximation to describe the electronic
properties of the poly-acetylene. But a more concise description requires
however taking into account the effects beyond the first nearest couplings.
These correlations can be thought of as corrections described by operators
of the form,%
\begin{equation}
\begin{tabular}{llll}
$\mathcal{F}_{nv}=\dsum\limits_{m\in \mathbb{Z}}A_{x_{m}}^{-}B_{x_{m}+nv}^{+}
$ & , & $\mathcal{F}_{nv}^{\dagger }=\dsum\limits_{m\in \mathbb{Z}%
}A_{x_{m}}^{+}B_{x_{m}+nv}^{-}$ & , \\
$\mathcal{G}_{nv}=\dsum\limits_{m\in \mathbb{Z}}A_{x_{m}}^{-}B_{x_{m}-nv}^{+}
$ & , & $\mathcal{G}_{nv}^{\dagger }=\dsum\limits_{m\in \mathbb{Z}%
}A_{x_{m}}^{+}B_{x_{m}-nv}^{-}$ & . \\
&  &  &
\end{tabular}%
\end{equation}%
with $n\geq 2$ generating an infinite dimensional Lie algebra. It happens
that the restriction to first nearest neighbors breaks this infinite
dimensional symmetry down to a sub-symmetry to be identified later on; see
\emph{property 3}.

\emph{property 2}: the structure of the operator $\mathcal{H}_{SU\left(
2\right) }$\ is very particular; it recalls basic features of complex
geometry, supersymmetry and step operators of Lie algebras. For the links
with complex geometry and supersymmetry, one has just to notice that given
some holomorphic function $F\left( Z\right) $ with a complex variable $Z$
that can be interpreted as a chiral superfield in 4D supersymmetry \textrm{%
\cite{bw}}; one may build two fundamental kinds of real quantities namely%
\begin{equation}
\begin{tabular}{llll}
$\left\vert F\left( Z\right) \right\vert ^{2}$ & \ or & $F\left( Z\right) +%
\bar{F}\left( \bar{Z}\right) $ & , \\
&  &  &
\end{tabular}%
\end{equation}%
in a quite similar manner with the Kahler $K\left( Z,\bar{Z}\right) $ and
chiral superpotentials $W\left( Z\right) $ of 4D supersymmetric non linear
sigma models. From this view, the hamiltonian $\mathcal{H}_{SU\left(
2\right) }$ is of the second type as it follows by setting
\begin{equation}
F\left( Z\right) =-t\left( \mathcal{F}_{v}+\mathcal{G}_{v}\right) ,
\end{equation}%
this feature may be also viewed as the reason behind the non diagonal form
of the hamiltonian.

\emph{property 3}: the third feature that we want to give here concerns the
relation between $\mathcal{H}_{SU\left( 2\right) }$ and $SU\left( 2\right) $
Lie algebra; as this result is also valid for higher dimensions and higher
rank groups; let us discuss it with some details here; and give just results
for higher dimensional lattices to be considered later on. \newline
There are two types of links between $\mathcal{H}_{SU\left( 2\right) }$ (\ref%
{23}); and the underlying $SU\left( 2\right) $ symmetry:

\begin{itemize}
\item the first link concerns the 1D lattice $\mathcal{L}_{su\left( 2\right)
}$ which , up to some details, is nothing but the weight lattice of the $%
SU\left( 2\right) $ Lie algebra. This is an integral lattice generated by
the fundamental weight $\mu _{0}$; i.e%
\begin{eqnarray}
\mathcal{L}_{su\left( 2\right) } &=&\left\{ x_{m}=mv=2dm\mu _{0},\quad m\in
\mathbb{Z}\right\} . \\
&&  \notag
\end{eqnarray}%
Notice that the sublattice $\mathcal{A}_{su\left( 2\right) }$ and $\mathcal{B%
}_{su\left( 2\right) }$ are related to the root lattice of $SU\left(
2\right) $ as already noticed before eq(\ref{R}); and the superposition is
done up to $\mu _{0}$ shifts. This property become clearer when we consider
bigger groups like SU$\left( 3\right) $; see for instance eqs(\ref{AA}-\ref%
{AB}) to fix the ideas.

\item the second link concerns the physics described by the hamiltonian $%
\mathcal{H}_{SU\left( 2\right) }$. From Lie algebra view, this hamiltonian
is a very special quantity given by the sum over step operators of some
infinite dimensional Lie algebra containing the following isomorphic $%
SU_{L}\left( 2\right) $ and $SU_{R}\left( 2\right) $ copies as subalgebras%
\begin{equation}
\begin{tabular}{llll}
$\left[ \mathcal{F}_{v},\mathcal{F}_{v}^{\dagger }\right] $ & $=$ & $%
\mathcal{N}$ &  \\
$\left[ \mathcal{N},\mathcal{F}_{v}\right] $ & $=$ & $-2\mathcal{F}_{v}$ &
\\
$\left[ \mathcal{N},\mathcal{F}_{v}^{\dagger }\right] $ & $=$ & $+2\mathcal{F%
}_{v}$ &
\end{tabular}%
\end{equation}%
and%
\begin{equation}
\begin{tabular}{llll}
$\left[ \mathcal{G}_{v},\mathcal{G}_{v}^{\dagger }\right] $ & $=$ & $%
\mathcal{N}^{\prime }$ &  \\
$\left[ \mathcal{N}^{\prime },\mathcal{G}_{v}\right] $ & $=$ & $-2\mathcal{G}%
_{v}$ &  \\
$\left[ \mathcal{N}^{\prime },\mathcal{G}_{v}^{\dagger }\right] $ & $=$ & $+2%
\mathcal{G}_{v}^{\dagger }$ &
\end{tabular}%
\end{equation}%
where%
\begin{equation}
\begin{tabular}{lll}
$\mathcal{N=}\dsum\limits_{m\in \mathbb{Z}}\left(
A_{x_{m}}^{+}A_{x_{m}}^{-}+B_{y_{m}}^{-}B_{y_{m}}^{+}\right) $ &  &  \\
&  &  \\
$\mathcal{N}^{\prime }=\dsum\limits_{m\in \mathbb{Z}}\left(
A_{x_{m}}^{+}A_{x_{m}}^{-}+B_{y_{m}^{\prime }}^{-}B_{y_{m}^{\prime
}}^{+}\right) $ &  &
\end{tabular}%
\end{equation}%
with $y_{m}=x_{m}+v$ and $y_{m}^{\prime }=x_{m}-v$.
\end{itemize}

\ \ \newline
To derive these commutation relations, we need to perform \emph{3} steps;
first use the realization eqs(\ref{FG}); second use also the algebra of the
fermionic operators $A_{x_{m}}^{\pm }$ and $B_{y_{m}}^{\pm }$satisfying the
usual anticommutation relations namely%
\begin{equation}
\begin{tabular}{llll}
$\left\{ A_{x_{m}}^{-},A_{x_{n}}^{+}\right\} =\delta _{nm}$ & , & $\left\{
A_{x_{m}}^{\pm },A_{x_{n}}^{\pm }\right\} =0$ & , \\
$\left\{ B_{y_{m}}^{-},B_{y_{n}}^{+}\right\} =\delta _{nm}$ & , & $\left\{
B_{y_{m}}^{\pm },B_{y_{n}}^{\pm }\right\} =0$ & ,%
\end{tabular}%
\end{equation}%
and
\begin{equation}
\begin{tabular}{ll}
$\left\{ A_{x_{m}}^{-},B_{y_{n}}^{\pm }\right\} =0$ &  \\
$\left\{ A_{x_{m}}^{+},B_{y_{n}}^{\pm }\right\} =0$ &
\end{tabular}%
\end{equation}%
and finally use the following relations
\begin{equation}
\begin{tabular}{llll}
$\left[ \mathcal{N},A_{x_{m}}^{+}\right] =+A_{x_{m}}^{+}$ & , & $\left[
\mathcal{N},A_{x_{m}}^{-}\right] =-A_{x_{m}}^{-}$ &  \\
$\left[ \mathcal{N},B_{y_{m}}^{+}\right] =-B_{y_{m}}^{+}$ & , & $\left[
\mathcal{N},B_{y_{m}}^{-}\right] =+B_{y_{m}}^{-}$ &
\end{tabular}%
\end{equation}%
that show, amongst others that, $\left( A_{\mathbf{x}_{m}}^{+},B_{y_{m}}^{+}%
\right) $ and $\left( B_{y_{m}}^{-},A_{y_{m}}^{-}\right) $ form $SU\left(
2\right) $ doublets under charge operator $\mathcal{N}$ and similarly with $%
\mathcal{N}^{\prime }$.

\ \ \ \newline
Performing the Fourier transform of $A_{x_{m}}^{\pm }$ and $%
B_{x_{m}+v_{i}}^{\pm }$ as given above, we can bring this hamiltonian to the
non diagonal form $\mathcal{H}_{SU\left( 2\right) }=\sum_{k}\mathcal{H}%
_{k}^{su_{2}}$ with
\begin{eqnarray}
&&  \notag \\
\mathcal{H}_{k}^{su_{2}} &=&\left( \tilde{A}_{k}^{-},\tilde{B}%
_{k}^{-}\right) \left(
\begin{array}{cc}
0 & e^{i2kd\mu _{0}}+e^{i2kd\mu _{1}} \\
e^{-i2kd\mu _{0}}+e^{-i2kd\mu _{1}} & 0%
\end{array}%
\right) \left(
\begin{array}{c}
\tilde{A}_{k}^{+} \\
\tilde{B}_{k}^{+}%
\end{array}%
\right)  \\
&&  \notag
\end{eqnarray}%
The diagonalization of this hamiltonian leads to the dispersion energy
relation,
\begin{equation}
E_{su\left( 2\right) }^{\pm }\left( k\right) =\pm t\sqrt{2+2\cos \left(
2kd\right) }
\end{equation}%
which is associated with the usual conducting band (+) and the valence one
(-). These relations can be also put into the form $\pm 2t\cos \left(
kd\right) $ from which we read that their zeros (Fermi energy) take place
for the wave vectors $k_{n}=\pm \frac{\pi }{2d}$ $\func{mod}$ $\frac{2\pi }{d%
}$.

\section{Graphene as a $SU(3)$ model}

Graphene, a sheet of graphite, is a 2D organic material system with carbons
in the $sp^{{\small 2}}$\ hybridization. This material, which is of great
interest nowadays, is expected to play a central role in nanotechnology
\textrm{\cite{9,10}}. The graphene lattice denoted here as $\mathcal{L}%
_{su\left( 3\right) }$ is a $2D$ honeycomb made by the superposition of two
triangular sublattices $\mathcal{A}_{su\left( 3\right) }$ and $\mathcal{B}%
_{su\left( 3\right) }$ as depicted in fig(2).

\begin{figure}[tbph]
\begin{center}
\hspace{0cm} \includegraphics[width=6cm]{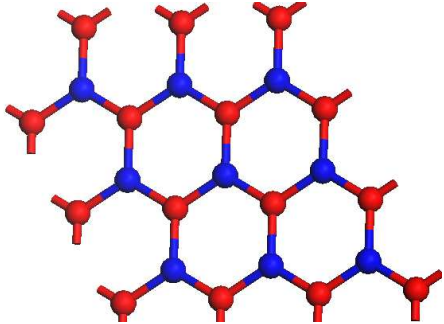}
\end{center}
\par
\vspace{-1cm}
\caption{{\protect\small Sublattices }$\mathcal{A}_{su\left( 3\right) }$%
{\protect\small \ (in blue) and }$\mathcal{B}_{su\left( 3\right) }$%
{\protect\small \ (in red) of the honeycomb. }}
\label{2}
\end{figure}

\ \ \ \ \newline
Each carbon atom, say $A_{\mathbf{r}_{n}}$ of the sublattice $\mathcal{A}%
_{su\left( 3\right) }$, has \emph{3} first nearest atom neighbors of $%
\mathcal{B}$-type namely $B_{\mathbf{r}_{n}+\mathbf{v}_{0}},$ $B_{\mathbf{r}%
_{n}+\mathbf{v}_{1}},B_{\mathbf{r}_{n}+\mathbf{v}_{2}}$ where $\mathbf{v}%
_{0},$\textbf{\ }$\mathbf{v}_{1},$\textbf{\ }$\mathbf{v}_{2}$ are \emph{2D}
vectors parameterizing their relative positions with respect to $A_{\mathbf{r%
}_{n}}$. These vectors satisfy the following constraint relation
\begin{equation}
\begin{tabular}{ll}
$\mathbf{v}_{0}+\mathbf{v}_{1}+\mathbf{v}_{2}=0$ & ,%
\end{tabular}
\label{BT3}
\end{equation}%
that should be compared with (\ref{BT2}). Like in the previous $SU\left(
2\right) $ case, this constraint equation turns out to have an
interpretation in terms of $SU\left( 3\right) $ group representations.
Setting $\mathbf{v}_{0}=\mathbf{\mu }_{0}d$, $\mathbf{v}_{1}=\mathbf{\mu }%
_{1}d,$ $\mathbf{v}_{2}=\mathbf{\mu }_{2}d$ where $d\simeq 1.42\mathring{A}$
is the length of the carbon-carbon bond, we end with the identity $\mathbf{%
\mu }_{0}+\mathbf{\mu }_{1}+\mathbf{\mu }_{2}=0$ describing precisely the
weight vectors of the $SU\left( 3\right) $ fundamental representation. To
fix the ideas on these weight vectors and the way they may be handled, we
give below two comments. \newline
(\textbf{1}) the explicit expressions of these weight vectors are given by
\begin{equation}
\begin{tabular}{llllll}
$\mathbf{\mu }_{1}=(\frac{\sqrt{2}}{2},\frac{\sqrt{6}}{6})$ & , & $\mathbf{%
\mu }_{2}=(-\frac{\sqrt{2}}{2},\frac{\sqrt{6}}{6})$ & , & $\mathbf{\mu }%
_{0}=-(0,\frac{\sqrt{6}}{3})$ & .%
\end{tabular}%
\end{equation}%
These planar vectors have the same norm $\left\Vert \mathbf{\mu }%
_{l}\right\Vert =\frac{2}{3}$ and the same angle $\left( \mathbf{\mu }_{i},%
\mathbf{\mu }_{j}\right) =\frac{2\pi }{3}$. They satisfy manifestly the
identity $\mathbf{\mu }_{0}+\mathbf{\mu }_{1}+\mathbf{\mu }_{2}=0$ capturing
the traceless property of the $SU\left( 3\right) $ symmetry. \newline
(\textbf{2}) the two generators of the sublattice $\mathcal{A}_{su\left(
3\right) }$ are given by the two simple roots $\mathbf{\alpha }_{1}$ and $%
\mathbf{\alpha }_{2}$ of the Lie algebra of the $SU\left( 3\right) $. These
two simple roots may be expressed in, different, but equivalent ways; one
way to do is in terms of the weight vectors $\mathbf{\mu }_{0},$ $\mathbf{%
\mu }_{1},$ $\mathbf{\mu }_{2}$ of the 3-dimensional representation of $%
SU\left( 3\right) $,
\begin{equation}
\mathbf{\alpha }_{1}=\mathbf{\mu }_{1}-\mathbf{\mu }_{2}\qquad ,\qquad
\mathbf{\alpha }_{2}=\mathbf{\mu }_{2}-\mathbf{\mu }_{0}.  \label{AA}
\end{equation}%
Another way is in terms of the two fundamental weight vectors $\mathbf{%
\omega }_{1}$ and $\mathbf{\omega }_{2}$ of $SU\left( 3\right) $ namely
\begin{equation*}
\mathbf{\alpha }_{1}=2\mathbf{\omega }_{1}-\mathbf{\omega }_{2}\qquad
,\qquad \mathbf{\alpha }_{2}=2\mathbf{\omega }_{2}-\mathbf{\omega }_{1}
\end{equation*}%
from which we learn
\begin{equation}
\begin{tabular}{lll}
$\mathbf{\mu }_{1}$ & $=\mathbf{\omega }_{1}$ & , \\
$\mathbf{\mu }_{2}$ & $=\mathbf{\omega }_{2}-\mathbf{\omega }_{1}$ & , \\
$\mathbf{\mu }_{0}$ & $=-\mathbf{\omega }_{2}$ & ,%
\end{tabular}%
\end{equation}%
and%
\begin{equation}
\mathbf{\alpha }_{i}.\mathbf{\omega }_{j}=\delta _{ij}.  \label{AB}
\end{equation}%
As such, sites $\mathbf{r}_{\mathbf{n}}^{A}$ and $\mathbf{r}_{\mathbf{n}}^{B}
$ in the sublattice $\mathcal{A}_{su\left( 3\right) }$ and $\mathcal{B}%
_{su\left( 3\right) }$ read respectively as follows,
\begin{equation}
\begin{tabular}{llll}
$\mathbf{r}_{\mathbf{n}}^{A}=n_{1}\mathbf{\alpha }_{1}+n_{2}\mathbf{\alpha }%
_{2}$ & , & $\mathbf{r}_{\mathbf{n}}^{B}=\mathbf{r}_{\mathbf{n}}^{A}+\mathbf{%
v}_{l}$ & ,%
\end{tabular}
\label{S}
\end{equation}%
where $\mathbf{n}=\left( n_{1}\mathbf{,}n_{2}\right) $ with $n_{1}$\textbf{\
}and\textbf{\ }$n_{2}$ integers; and where the \emph{3} $\mathbf{v}_{l}$'s
are as before.\newline
The tight binding hamiltonian $\mathcal{H}_{SU\left( 3\right) }$ describing
the electronic correlations restricted to the first nearest neighbors is
given by
\begin{equation}
\begin{tabular}{lll}
$\mathcal{H}_{SU\left( 3\right) }$ & $=-t\left( \mathcal{F}_{\mathbf{v}_{1}}+%
\mathcal{F}_{\mathbf{v}_{1}}^{\dagger }\right) -t\left( \mathcal{F}_{\mathbf{%
v}_{2}}+\mathcal{F}_{\mathbf{v}_{2}}^{\dagger }\right) -t\left( \mathcal{F}_{%
\mathbf{v}_{0}}+\mathcal{F}_{\mathbf{v}_{0}}^{\dagger }\right) $ & ,%
\end{tabular}%
\end{equation}%
where $t$ is the hop energy and where $\mathcal{F}_{\mathbf{v}_{l}}$ are
step operators given by%
\begin{equation}
\begin{tabular}{lll}
$\mathcal{F}_{\mathbf{v}_{l}}$ & $=\dsum\limits_{m}A_{\mathbf{r}_{m}}^{-}B_{%
\mathbf{r}_{m}+\mathbf{v}_{l}}^{+}$ & ,%
\end{tabular}%
\end{equation}%
with $A_{\mathbf{r}_{m}}^{\pm }$ and $B_{\mathbf{r}_{m}+\mathbf{v}_{l}}^{\pm
}$ satisfying the electronic anticommutation relations. Notice that the
hamiltonian $\mathcal{H}_{SU\left( 3\right) }$ and the \emph{3} step
operators $\mathcal{F}_{\mathbf{v}_{l}}$ are quite similar to those of the
poly-acetylene chain namely $\mathcal{H}_{SU\left( 2\right) }$ and the
associated $\mathcal{F}_{v}$ and $\mathcal{G}_{v}$ operators. From these $%
\mathcal{F}_{\mathbf{v}_{l}}$'s, we can make \emph{3} copies of triplets%
\begin{equation}
\begin{tabular}{llll}
$\mathcal{F}_{\mathbf{v}_{l}},$ & $\mathcal{F}_{\mathbf{v}_{l}}^{\dagger },$
& $\mathcal{N}_{\mathbf{v}_{l}}=\left[ \mathcal{F}_{\mathbf{v}_{l}},\mathcal{%
F}_{\mathbf{v}_{l}}^{\dagger }\right] $ & ,%
\end{tabular}%
\end{equation}%
each copy obeys an \emph{SU}$\left( 2\right) $ Lie algebra. Notice also that
the $\mathcal{F}_{\mathbf{v}_{l}}$'s and $\mathcal{F}_{\mathbf{v}%
_{l}}^{\dagger }$'s are in fact particular operators of more general ones
describing generic hops and generating an infinite dimensional symmetry.

\ \ \ \newline
Performing the Fourier transform of these field operators, we can put $%
\mathcal{H}_{SU\left( 3\right) }$ into the form $\sum_{\mathbf{k}}\mathcal{H}%
_{\mathbf{k}}^{su_{3}}$ where $\mathbf{k}=\left( k_{x},k_{y}\right) $ is the
wave vector and where
\begin{equation}
\begin{tabular}{ll}
$\mathcal{H}_{\mathbf{k}}^{su_{3}}=$ & $\left( \tilde{A}_{\mathbf{k}}^{-},%
\tilde{B}_{\mathbf{k}}^{-}\right) \left(
\begin{array}{cc}
0 & \varepsilon _{su_{3}}\left( \mathbf{k}\right)  \\
\overline{\varepsilon _{su_{3}}\left( \mathbf{k}\right) } & 0%
\end{array}%
\right) \left(
\begin{array}{c}
\tilde{A}_{\mathbf{k}}^{+} \\
\tilde{B}_{\mathbf{k}}^{+}%
\end{array}%
\right) $%
\end{tabular}%
\end{equation}%
with%
\begin{equation}
\begin{tabular}{ll}
$\varepsilon _{su_{3}}\left( \mathbf{k}\right) =e^{id\mathbf{k.\mu }%
_{0}}+e^{id\mathbf{k.\mu }_{1}}+e^{id\mathbf{k.\mu }_{2}}$ & .%
\end{tabular}%
\end{equation}%
The diagonalization of this hamiltonian leads to the dispersion energy
relations%
\begin{equation}
E_{su\left( 3\right) }^{\pm }\left( \mathbf{k}\right) =\pm t\sqrt{3+2\left[
\cos \left( d\mathbf{k.\alpha }_{1}\right) +\cos \left( d\mathbf{k.\alpha }%
_{2}\right) +\cos \left( d\mathbf{k.\alpha }_{3}\right) \right] },
\end{equation}%
with
\begin{equation}
\begin{tabular}{llllll}
$\mathbf{\alpha }_{1}=\mathbf{\mu }_{1}-\mathbf{\mu }_{2}$ & , & $\mathbf{%
\alpha }_{2}=\mathbf{\mu }_{2}\mathbf{-\mu }_{0}$ & , & $\mathbf{\mu }_{0}-%
\mathbf{\mu }_{1}=\mathbf{\alpha }_{3}=\mathbf{\alpha }_{1}+\mathbf{\alpha }%
_{2}$ & ,%
\end{tabular}%
\end{equation}%
nothing but the positive roots of the $SU\left( 3\right) $ symmetry. We will
show below that the apparition of the roots in the dispersion energies is a
general feature of tight binding model shared by the series of models
classified by the Lie groups.

\section{$SU\left( 4\right) $ diamond model}

\begin{figure}[tbph]
\begin{center}
\hspace{0cm} \includegraphics[width=6cm]{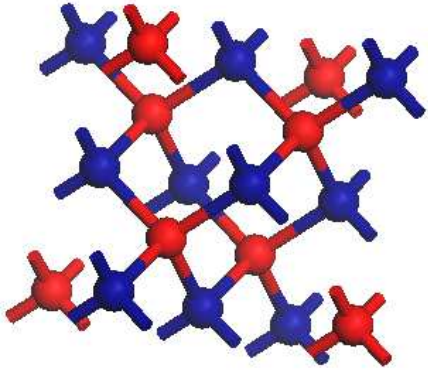}
\end{center}
\par
\vspace{-1cm}
\caption{the lattice $\mathcal{L}_{su\left( 4\right) }$ with {\protect\small %
sublattices }$\mathcal{A}_{su\left( 4\right) }${\protect\small \ (in blue)
and }$\mathcal{B}_{su\left( 4\right) }${\protect\small \ (in red). Each atom
has \emph{4} first nearest neighbors forming a tetrahedron and \emph{12}
second nearest ones.}}
\label{3}
\end{figure}

In this model, the lattice $\mathcal{L}_{su\left( 4\right) }$ is a
3-dimensional crystal made by two tetrahedral sublattices $\mathcal{A}%
_{su\left( 4\right) }$ and $\mathcal{B}_{su\left( 4\right) }$ whose
superposition follows the same logic as in the poly-acetylene chain and the
2D honeycomb; see eqs(\ref{S}) with $\mathbf{r}_{\mathbf{n}}^{A}=n_{1}%
\mathbf{\alpha }_{1}+n_{2}\mathbf{\alpha }_{2}+n_{3}\mathbf{\alpha }_{3}$,
the relative vectors $\mathbf{v}_{l}$ as in (\ref{NB})-(\ref{vi}); and fig(%
\ref{3}) for illustration.

\ \ \ \ \newline
This physics describing electronic properties by using tight binding
approach requires to go beyond the $sp^{1}$ and $sp^{2}$ hybridizations of
the carbon atom considered in the two previous examples. Here, each atom,
say of type $A_{\mathbf{r}_{\mathbf{n}}}$ located at site $\mathbf{r}_{%
\mathbf{n}},$ has \emph{4} first nearest neighbors that can make 4 localized
$\sigma $- bonds,
\begin{equation}
A_{\mathbf{r}_{\mathbf{n}}}\rightarrow \left\{
\begin{array}{c}
B_{\mathbf{r}_{\mathbf{n}}+\mathbf{v}_{0}} \\
B_{\mathbf{r}_{\mathbf{n}}+\mathbf{v}_{1}} \\
B_{\mathbf{r}_{\mathbf{n}}+\mathbf{v}_{2}} \\
B_{\mathbf{r}_{\mathbf{n}}+\mathbf{v}_{3}}%
\end{array}%
\right.  \label{NB}
\end{equation}%
So in case where atoms are carbons, there is no delocalized electrons of
type $\pi $ that can hop from a carbon to its closest neighbors; this is why
carbon-diamond is an insulator. \newline
\textrm{Nevertheless, in case where we have atoms that allows }$sp^{3}d^{1}$%
\textrm{\ }with delocalized electrons hoping to nearest neighboring
tetrahedral sites; the tight binding approach applies as in previous case.
To our understanding, this concerns in general material alloys with $%
sp^{3}d^{n}$ hybridizations involving certain atoms of transition metals
with delocalized d-electrons; for a related $sp^{3}$ tight-binding study
concerning the calculation of the electronic and optical properties see
\textrm{\cite{SP}}.\newline
On the other hand, as far as tight binding model on \emph{3D} diamond is
concerned, one also may engineer simple toy models to describe quite similar
physical properties as the electronic ones. Below, we develop a lattice
model describing dynamical vacancies within the \emph{3D} crystal,%
\begin{equation}
\mathcal{L}_{su\left( 4\right) }=\mathcal{A}_{su\left( 4\right) }\cup
\mathcal{B}_{su\left( 4\right) }
\end{equation}%
This is an ideal and simple toy model that is based on the two following
heuristic hypothesizes:

\begin{itemize}
\item we assume that in the 3D lattice in its fundamental state has sites of
the sublattice $\mathcal{A}_{su\left( 4\right) }$ occupied by atoms; and the
site of $\mathcal{B}_{su\left( 4\right) }$ are vacancies; that is are
unoccupied free sites.

\item we also assume that under some external parameters, the atoms of $%
\mathcal{A}_{su\left( 4\right) }$ leave their initial positions towards the
free nearest neighbors in $\mathcal{B}_{su\left( 4\right) }$; that A-atoms
jump to $\mathcal{B}_{su\left( 4\right) }$ and B-vacancies jump to $\mathcal{%
A}_{su\left( 4\right) }$.
\end{itemize}

\ \newline
Actually one needs to specify other data; but let us forget about them by
focusing on the ideal configuration; and look for the dispersion energies of
the hop of atoms and vacancies in the \emph{3D} lattice.\newline
To that purpose, notice first that each site $\mathbf{r}_{m}$ in $\mathcal{L}%
_{su\left( 4\right) }$ has \emph{4} first nearest neighbors at $\left(
\mathbf{r}_{m}+\mathbf{v}_{i}\right) $ as in eq(\ref{NB}) forming the
vertices of a regular tetrahedron. These relative positions $\mathbf{v}_{i}$
are given by:%
\begin{equation}
\begin{tabular}{llll}
$\mathbf{v}_{1}=\frac{d}{\sqrt{3}}\left( -1,-1,+1\right) $ & , & $\mathbf{v}%
_{2}=\frac{d}{\sqrt{3}}\left( -1,+1,-1\right) $ &  \\
$\mathbf{v}_{3}=\frac{d}{\sqrt{3}}\left( +1,-1,-1\right) $ & , & $\mathbf{v}%
_{0}=\frac{d}{\sqrt{3}}\left( +1,+1,+1\right) $ &
\end{tabular}
\label{vi}
\end{equation}%
and obey the constraint relation
\begin{equation*}
\mathbf{v}_{1}+\mathbf{v}_{2}+\mathbf{v}_{3}+\mathbf{v}_{0}=0
\end{equation*}%
that should be compared with (\ref{BT2},\ref{BT3}). Like in the previous $%
SU\left( 2\right) $ and $SU\left( 3\right) $ cases, this constraint equation
has an interpretation in terms of $SU\left( 4\right) $ group. Setting $%
\mathbf{v}_{i}=\frac{d}{\sqrt{3}}\mathbf{\mu }_{i}$, we end with
\begin{equation}
\begin{tabular}{ll}
$\mathbf{\mu }_{0}+\mathbf{\mu }_{1}+\mathbf{\mu }_{2}+\mathbf{\mu }_{3}=%
\mathbf{0}$ & ,%
\end{tabular}
\label{m}
\end{equation}%
showing that the $\mathbf{\mu }_{i}$'s are the weight vectors of the
fundamental representation of SU$\left( 4\right) $ with dominant weight $%
\mathbf{\mu }_{0}$. \newline
Let $A_{\mathbf{r}_{m}}$ and $B_{\mathbf{r}_{m}+\mathbf{v}_{i}}$ be the
quantum states describing the particle at $\mathbf{r}_{m}$ and the vacancy
at $\mathbf{r}_{m}+\mathbf{v}_{i}$ respectively. Let also $A_{\mathbf{r}%
_{m}}^{\pm }$ and $B_{\mathbf{r}_{m}+\mu _{i}}^{\pm }$ be the corresponding
creation and annihilation operators. The hamiltonian describing the hop of
the vacancy/particle to the first nearest neighbors is given by
\begin{equation}
\begin{tabular}{lll}
$\mathcal{H}_{SU\left( 4\right) }=$ & $-t\left( \mathcal{F}_{\mathbf{v}_{1}}+%
\mathcal{F}_{\mathbf{v}_{1}}^{\dagger }\right) -t\left( \mathcal{F}_{\mathbf{%
v}_{2}}+\mathcal{F}_{\mathbf{v}_{2}}^{\dagger }\right) $ &  \\
& $-t\left( \mathcal{F}_{\mathbf{v}_{3}}+\mathcal{F}_{\mathbf{v}%
_{3}}^{\dagger }\right) -t\left( \mathcal{F}_{\mathbf{v}_{0}}+\mathcal{F}_{%
\mathbf{v}_{0}}^{\dagger }\right) $ & ,%
\end{tabular}%
\end{equation}%
where the $\mathcal{F}_{\mathbf{v}_{l}}$'s are step operators given by%
\begin{equation}
\begin{tabular}{lll}
$\mathcal{F}_{\mathbf{v}_{l}}$ & $=\dsum\limits_{m}A_{\mathbf{r}_{m}}^{-}B_{%
\mathbf{r}_{m}+\mathbf{v}_{l}}^{+}$ & .%
\end{tabular}%
\end{equation}
By performing Fourier transforms of the $A_{\mathbf{r}_{m}}^{\pm }$, $B_{%
\mathbf{r}_{m}+\mu _{i}}^{\pm }$ field operators and following the same
steps as in the SU$\left( 2\right) $ and SU$\left( 3\right) $ cases, we end
with the remarkable form of the dispersion energies%
\begin{equation}
\begin{tabular}{ll}
$E_{su_{4}}^{\pm }\left( \mathbf{k}\right) =\pm t\sqrt{4+2\dsum\limits_{0%
\leq i<j\leq 3}\cos \left( \frac{d}{\sqrt{3}}\mathbf{k.\alpha }_{ij}\right) }
$ & ,%
\end{tabular}%
\end{equation}%
with $\mathbf{k}=\left( k_{x},k_{y},k_{z}\right) $ is the wave vector and
where%
\begin{equation}
\begin{tabular}{ll}
$\mathbf{\alpha }_{ij}\mathbf{=\mu }_{i}-\mathbf{\mu }_{j}$ & ,%
\end{tabular}%
\end{equation}%
are exactly the \emph{12} roots of the $SU\left( 4\right) $ group.

\section{ Octahedral $SO\left( 6\right) $ model}

\begin{figure}[tbph]
\begin{center}
\hspace{0cm} \includegraphics[width=4cm]{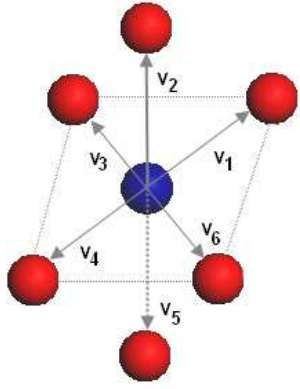}
\end{center}
\par
\vspace{-1cm}
\caption{E{\protect\small ach atom has \emph{6} first nearest forming an
octahedron}$.$}
\label{br}
\end{figure}

The lattice $\mathcal{L}_{SO\left( 6\right) }$, whose unit cell is given by
the figure (\ref{br}), is an other 3-dimensional crystal made by the
superposition of two octahedral sublattices $\mathcal{A}_{SO\left( 6\right)
} $ and $\mathcal{B}_{SO\left( 6\right) }$. Each atom $A_{\mathbf{r}_{%
\mathbf{n}}}$ located at site $\mathbf{r}_{\mathbf{n}}$ has \emph{6} first
nearest neighbors,
\begin{equation}
A_{\mathbf{r}_{\mathbf{n}}}\rightarrow \left\{
\begin{array}{c}
B_{\mathbf{r}_{\mathbf{n}}+\mathbf{v}_{1}} \\
B_{\mathbf{r}_{\mathbf{n}}+\mathbf{v}_{2}} \\
B_{\mathbf{r}_{\mathbf{n}}+\mathbf{v}_{3}} \\
B_{\mathbf{r}_{\mathbf{n}}+\mathbf{v}_{4}} \\
B_{\mathbf{r}_{\mathbf{n}}+\mathbf{v}_{5}} \\
B_{\mathbf{r}_{\mathbf{n}}+\mathbf{v}_{6}}%
\end{array}%
\right.  \label{RB}
\end{equation}%
Unlike the $SU\left( 4\right) $ model, each atom in the lattice $\mathcal{L}%
_{SO\left( 6\right) }$ give rather \emph{6} bonds as in fig(\ref{br}); so
this model might be used to describe the electronic properties of material
alloys which have $sp^{3}d^{2}$ hybridization with delocalized d-electrons
hoping to nearest neighboring octahedral sites.\newline
Below, we will also use the toy model described in section 4; but now with
the \emph{3D} lattice $\mathcal{L}_{SO\left( 6\right) }$. Sites of the
sublattice $\mathcal{A}_{SO\left( 6\right) }$ are occupied by atoms and
those of the site $\mathcal{B}_{SO\left( 6\right) }$ are free vacancies. We
also assume that under external parameters, the atoms and vacancies of $%
\mathcal{L}_{SO\left( 6\right) }$ leave their initial positions and move
towards the first nearest neighbors.

To study the dynamical properties of this toy model, notice first that each
site $\mathbf{r}_{m}$ in $\mathcal{L}_{SO\left( 6\right) }$ has \emph{6}
first nearest neighbors at $\left( \mathbf{r}_{m}+\mathbf{v}_{I}\right) $
forming the vertices of an octahedron as shown on the fig(\ref{br}) and eq(%
\ref{RB}). Like in previous models, the relative positions $\mathbf{v}_{I}$
obey the constraint relation
\begin{equation}
\begin{tabular}{ll}
$\mathbf{v}_{1}+\mathbf{v}_{2}+\mathbf{v}_{3}+\mathbf{v}_{4}+\mathbf{v}_{5}+%
\mathbf{v}_{6}=0$ & ,%
\end{tabular}
\label{BT4}
\end{equation}%
having as well a group theoretic interpretation depending on the way it is
solved. Strictly speaking, there are various ways to solve this constraint
relation; but for the case at hand, it is solved as follows: First split the
set of the \emph{6} vectors into two subsets like,
\begin{equation}
\begin{tabular}{llll}
$\mathbf{v}_{I}$ & $=\left\{
\begin{array}{c}
v_{i}^{+}=\mathbf{v}_{i}\text{ \ \ } \\
v_{i}^{-}=\mathbf{v}_{3+i}%
\end{array}%
\right. $ & , & $i=1,2,3$%
\end{tabular}%
\end{equation}%
so that the constraint relation (\ref{BT4}) takes the form $\sum_{i}\left(
v_{i}^{+}+v_{i}^{-}\right) $. Then solve the vanishing condition by taking $%
v_{i}^{-}=-v_{i}^{+}$; this leads to the octahedron of fig(\ref{br}).
Moreover, setting $\mathbf{v}_{I}=d\mathbf{\mu }_{I}$ with factor $d$ the
atom-vacancy distance, one can check explicitly that the $\mathbf{\mu }_{I}$%
s are nothing but the weights of the $SO\left( 6\right) $ vector
representation. Recall that $SO\left( 6\right) $ has rank \emph{3} and
dimension \emph{15}; it also has 12 roots $\mathbf{\alpha }_{IJ}$, three of
them denoted $\mathbf{\alpha }_{i}$ are simple; the $\mathbf{\alpha }_{IJ}$s
are like $\pm \left( \mathbf{\mu }_{i}\pm \mathbf{\mu }_{j}\right) $ with $%
1\leq i,j\leq 3$ and the simple ones are given by
\begin{equation}
\begin{tabular}{lll}
$\mathbf{\alpha }_{1}=\mathbf{\mu }_{1}-\mathbf{\mu }_{2},$ & $\mathbf{%
\alpha }_{2}=\mathbf{\mu }_{2}-\mathbf{\mu }_{3},$ & $\mathbf{\alpha }_{3}=%
\mathbf{\mu }_{2}+\mathbf{\mu }_{3}.$%
\end{tabular}%
\end{equation}%
The tight binding hamiltonian $\mathcal{H}_{SO\left( 6\right) }$ describing
the hop of the vacancy/particle to the first nearest neighbors is given by,
\begin{equation}
\begin{tabular}{lll}
$\mathcal{H}_{SO\left( 6\right) }=$ & $-t\left( \mathcal{F}_{\mathbf{v}_{1}}+%
\mathcal{F}_{\mathbf{v}_{1}}^{\dagger }\right) -t\left( \mathcal{F}_{\mathbf{%
v}_{2}}+\mathcal{F}_{\mathbf{v}_{2}}^{\dagger }\right) -t\left( \mathcal{F}_{%
\mathbf{v}_{3}}+\mathcal{F}_{\mathbf{v}_{3}}^{\dagger }\right) $ &  \\
& $-t\left( \mathcal{F}_{\mathbf{v}_{4}}+\mathcal{F}_{\mathbf{v}%
_{4}}^{\dagger }\right) -t\left( \mathcal{F}_{\mathbf{v}_{5}}+\mathcal{F}_{%
\mathbf{v}_{5}}^{\dagger }\right) -t\left( \mathcal{F}_{\mathbf{v}_{6}}+%
\mathcal{F}_{\mathbf{v}_{6}}^{\dagger }\right) $ & ,%
\end{tabular}%
\end{equation}%
where the $\mathcal{F}_{\mathbf{v}_{l}}$'s are step operators given by $%
\sum_{m}A_{\mathbf{r}_{m}}^{-}B_{\mathbf{r}_{m}+\mathbf{v}_{l}}^{+}$. By
performing Fourier transforms of the $A_{\mathbf{r}_{m}}^{\pm }$, $B_{%
\mathbf{r}_{m}+_{V_{I}}}^{\pm }$ field operators and following the same
steps as before, we end with the dispersion energy relations%
\begin{equation}
\begin{tabular}{ll}
$E_{so_{6}}^{\pm }\left( \mathbf{k}\right) =\pm t\sqrt{6+2\dsum\limits_{0%
\leq I<J\leq 6}\cos \left( d\text{ }\mathbf{k.\alpha }_{IJ}\right) }$ & ,%
\end{tabular}%
\end{equation}%
with $\mathbf{k=}\left( k_{x},k_{y},k_{z}\right) $ is the wave vector and $%
\mathbf{\alpha }_{IJ}$ the roots of $SO\left( 6\right) $.

\section{Conclusion and comments}

\begin{figure}[tbph]
\begin{center}
\hspace{0cm} \includegraphics[width=6cm]{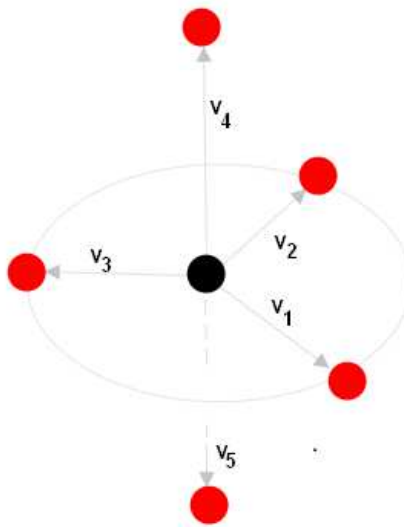}
\end{center}
\par
\vspace{-1cm}
\caption{\emph{bonds in the dsp}$^{3}$\emph{\ hybridization with the 5 first
nearest neighbors form a\ vector representation of a hidden SO}$\left( \emph{%
5}\right) $\emph{\ symmetry.}}
\label{SO5}
\end{figure}

In the present paper, we have used tight binding approach and continuous
group representation methods to engineer a series of lattice QFT models
classified by ADE Lie groups. These lattice models include the electronic
properties of the \emph{1D} poly-acetylene chain and \emph{2D} graphene, but
also dynamical properties of the vacancies in lattice models such as the $%
SU\left( 4\right) $ diamond and the $SO\left( 6\right) $ octahedron
considered in this study. As a remarkable fact following this set of models
is that the dispersion energies, restricted to first nearest neighbors
couplings, are completely characterized by the roots of the Lie algebras
underlying the crystals. The lattice QFT models considered here are
associated with low rank groups as required by condensed matter systems with
real dimension at most \emph{3}. If relaxing this condition to include
higher dimensions; the results obtained in this paper extend naturally to
the $SU\left( N\right) $ and $SO\left( N\right) $ series. We end this study
by first noting that there still is an interesting issue that might have
interpretation in condensed matter systems; it concerns the models based on $%
SO\left( 5\right) $ and $G_{2}$ groups having rank \emph{2} and associated
with non simply laced Lie algebras; atoms living at sites of the lattice $%
\mathcal{L}_{SO\left( 5\right) }$, whose \emph{5} first nearest neighbors
form a\ vector representation of a hidden $SO\left( 5\right) $ symmetry,
allows five bonds and would corresponds to the \emph{sp}$^{3}$\emph{d}$^{1}$
hybridization as shown of fig(\ref{SO5}).

Notice also that lattice model based on \emph{4D} hyperdiamond has $SU\left(
5\right) $ symmetry; it has been related in \textrm{\cite{55,11}} with
lattice QCD; see also \textrm{\cite{12}-\cite{14},\cite{56}}.

\begin{acknowledgement}
: L.B Drissi would like to thank the ICTP Associationship program for
support. E.H Saidi thanks URAC 09.
\end{acknowledgement}


\begin{thebibliography}{99}
\bibitem{1} P.R Wallace, Phys Rev 71, (1947), 622,

\bibitem{2} G.V. Semenoff, Phys. Rev. Lett. 53, 2449 (1984),

\bibitem{3} K. S Novoselov et al, Science 306 (2004) 666,

\bibitem{4} A.H.Castro-Neto et al. Rev. Mod. Phys.81, 109 (2009),

\bibitem{5} L.B Drissi, E.H Saidi, M.Bousmina, Nucl Phys B, Vol 829,
p.523-533, arXiv:1106.5222,

\bibitem{55} L.B Drissi, E.H Saidi, M.Bousmina, \emph{4D graphene}, Phys Rev
D (2011), in press, arXiv:1106.5222

\bibitem{56} L.B Drissi, H. Mhamdi, E.H Saidi, \emph{Anomalous Quantum Hall
Effect of 4D Graphene in Background Fields}, arXiv:1106.5578,

\bibitem{DS} M.S. Dresselhaus, G. Dresselhaus, A. Jorio, \emph{Group Theory.
Applications to the Physics of Condensed Matter}, Springer, Berlin, 2008,

\bibitem{6} Brian C. Hall Lie Groups, \emph{Lie Algebras, and
Representations: An Elementary Introduction,} Springer, 2003. ISBN
0-387-40122-9,

\bibitem{7} P. Goddard and D. Olive, \emph{Kac-Moody and Virasoro algebras
in relation to quantum physics,} Int. Journ. of Mod. Phys. A 1, 303-414,
1986,

\bibitem{8} R. Ahl Laamara, M. Ait Ben Haddou, A Belhaj, L.B Drissi, E.H
Saidi, Nucl.Phys. B702 (2004) 163-188,

\bibitem{9} Gunn Kim, Yeonju Kim, Jisoon Ihm, Chemical Physics Letters v.
415, p.279 (2005), arXiv:cond-mat/0506748,

\bibitem{bw} J. Wess, J. Bagger, \emph{Supersymmetry and Supergravity},
Princeton University Press, Prince- ton, 1983,

\bibitem{10} A. K. Geim, Graphene: \emph{Status and Prospects}, Science 19,
Vol. 324. no. 5934, pp. 1530 - 1534, (2009),

\bibitem{SP} S. Schulz, S. Schumacher, G. Czycholl, Phys. Rev. B73, 245327
(2006), arXiv:0802.2436,

\bibitem{11} Lalla Btissam Drissi, El Hassan Saidi, \emph{On Dirac Zero
Modes in Hyperdiamond Model}, arXiv:1103.1316, Under consideration in Phys
Rev D.

\bibitem{12} Michael Creutz, \emph{Four dimensional graphene and chiral
fermions}, JHEP0804:017,2008, arXiv:0712.1201,

\bibitem{13} Taro Kimura, Tatsuhiro Misumi, Prog.Theor.Phys.123: 63-78, $%
\left( 2010\right) $, arXiv:0907.3774,

\bibitem{14} T. Kimura and T. Misumi, Prog.Theor.Phys.124: 415-432, $\left(
2010\right) $, arXiv:0907.1371,

\bibitem{15} Michael Creutz, Tatsuhiro Misumi, \emph{Classification of
Minimally Doubled Fermions}, Phys.Rev.D82:074502,2010, arXiv:1007.3328,
\end{thebibliography}
\end{document}